# FUZZY FEATURE SELECTION WITH KEY-BASED CRYPTOGRAPHIC TRANSFORMATIONS

MIKE NKONGOLO
DEPARTMENT OF INFORMATICS
UNIVERSITY OF PRETORIA
PRIVATE BAG X20, HATFIELD, 0028
SOUTH AFRICA
Email: mike.wankongolo@up.ac.za

**Abstract.** In the field of cryptography, the selection of relevant features plays a crucial role in enhancing the security and efficiency of cryptographic algorithms. This paper presents a novel approach of applying fuzzy feature selection to key-based cryptographic transformations. The proposed fuzzy feature selection leverages the power of fuzzy logic to identify and select optimal subsets of features that contribute most effectively to the cryptographic transformation process. By incorporating fuzzy feature selection into key-based cryptographic transformations, this research aims to improve the resistance against attacks and enhance the overall performance of cryptographic systems. Experimental evaluations may demonstrate the effectiveness of the proposed approach in selecting secure key features with minimal computational overhead. This paper highlights the potential of fuzzy feature selection as a valuable tool in the design and optimization of key-based cryptographic algorithms, contributing to the advancement of secure information exchange and communication in various domains.

Fuzzy logic is a powerful mathematical framework that addresses uncertainty and imprecision in reasoning and decision-making [4]. Unlike traditional binary logic, which relies on a dichotomy of true and false, fuzzy logic allows for the representation of degrees of truth [3]. It introduces fuzzy sets, where elements can have membership values ranging between 0 and 1, reflecting their degree of belonging to a set [3, 4]. This capability enables modeling and reasoning with vague and ambiguous concepts, facilitating flexible and nuanced decision-making in complex and uncertain domains [5]. Fuzzy logic finds applications in diverse fields, including control systems, artificial intelligence, pattern recognition, and decision support systems [3, 4, 5, 8, 9, 10, 11, 12, 13, 14]. By incorporating fuzzy logic into the feature selection process, the proposed method enables the identification of relevant patterns that can contribute to more robust cryptographic transformations. The utilization of a dynamic fuzzy network allows for adaptability and responsiveness to changing network conditions, thereby enhancing the security posture of cryptographic systems. In the context of network security, feature selection plays a crucial role in identifying the most significant factors that contribute to the effectiveness of cryptographic transformations. By employing fuzzy logic and the dynamic fuzzy network, the proposed approach ensures that the selection of features is based on their relevance, thereby optimizing the security measures employed within the network. The object-oriented framework utilized in the implementation facilitates the efficient storage and processing of critical attributes, enabling seamless integration of the fuzzy logic approach into existing cryptographic systems. The incorporation of fuzzy logic in feature selection not only enhances the security capabilities of cryptographic transformations but also addresses the complexity and computational challenges involved. By identifying pertinent patterns and optimizing the selection of features, the proposed methodology minimizes computational overhead while maximizing security. This research contribution offers a novel perspective in the realm of cryptographic transformation for network security and opens up new avenues for further exploration and advancements in this critical domain. Section 2 introduces a method for implementing the feature selection step in dynamic fuzzy networks. This step involves propagating input values through the network layers, calculating membership functions at each node for the selected features, and passing the information to subsequent layers. Additionally, Section 3 presents a method for implementing the propagation technique to train these dynamic fuzzy networks. This technique allows for the adjustment of weights and biases based on a comparison between the network's output and the desired output. Lastly, Section 4 presents the application of key based cryptographic transformation to the selected fuzzy sets. It is presumed that the reader possesses a fundamental understanding of fuzzy networks. A comprehensive introduction to these structures can be referenced in established works such as [1], [2], and [3].

## Dynamic fuzzy networks

In the context of fuzzy logic for feature selection, we introduce the concept of a dynamic fuzzy network, which can be seen as a collection of fuzzy nodes (Figure 1). This network structure plays a crucial role in representing and selecting important features during the selection process. A dynamic fuzzy network consists of interconnected layer of nodes, forming a linked list [3, 4]. The input-layer node is positioned at the beginning, while the output-layer node is located at the end of the network [1]. Figure 1 visually illustrates the layer nodes as squares, representing the flow of information within the dynamic network. In the context of the proposed architecture, each layer is responsible for selecting relevant features. However, it is the input layer, specifically the node labeled as n, that records all similar patterns. This node encapsulates the membership function and other pertinent properties associated with specific features similarities. By leveraging the principles of fuzzy logic [3, 4], the dynamic nature of the proposed fuzzy network allows for the adaptation and refinement of the fuzzy nodes, facilitating feature selection based on fuzzy rules. The connectivity between the layers ensures the seamless integration of selected features throughout the dynamic network, enabling the smooth flow of information. In the context of fuzzy logic for feature selection, this depicted dynamic fuzzy network structure plays a pivotal role. Each node in the network represents a fuzzy set, capturing the inherent uncertainty and imprecision associated with features [3]. The interconnected layer nodes form a linked list, facilitating the flow of information from the input-layer node to the output-layer node [1]. During the feature selection process, the node **n** in the input layer nodes records all the relevant features that share similarities within each fuzzy set. This dynamic structure can be incorporated into the cryptographic framework to achieve effective feature selection based on fuzzy logic principles. The network's ability to adapt and modify the fuzzy nodes enhances the selection and refinement of features, contributing to improved cryptographic transformations. This research underscores the significance of fuzzy logic in feature selection within the cryptographic context and offers valuable insights for further advancements in this field.

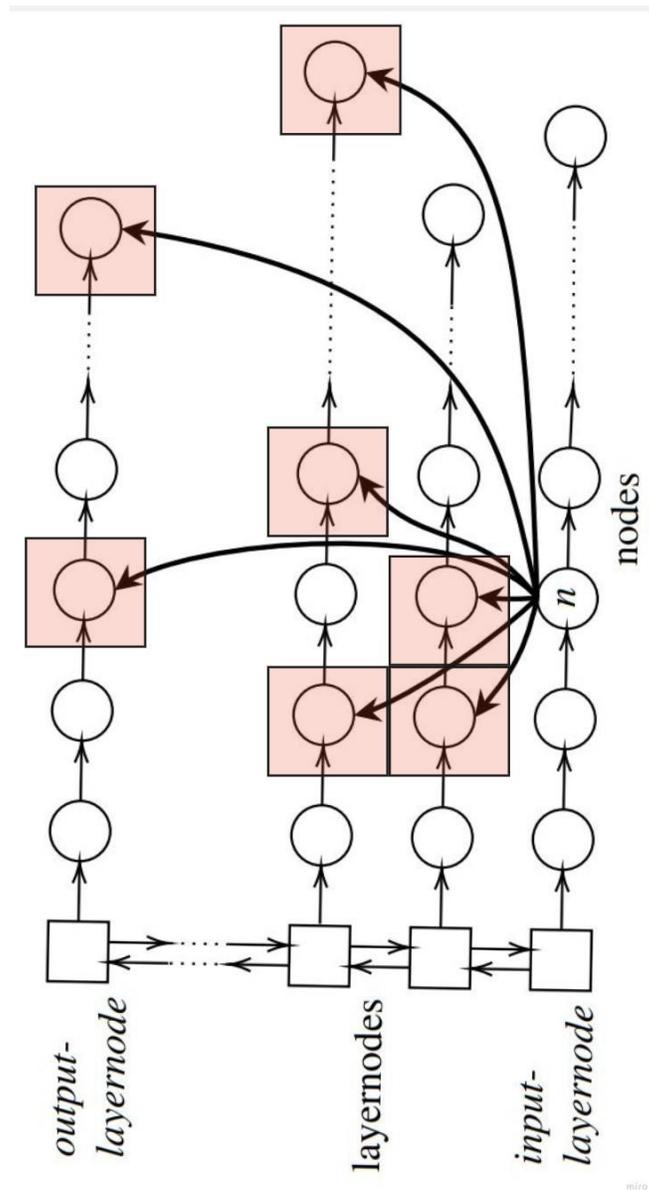

Figure 1. The structure of a dynamic fuzzy network for feature selection.

## Feature selection and dynamic fuzzy networks

The dynamic fuzzy logic for feature selection has been illustrated in Algorithm 1.

**Algorithm 1:** Dynamic Fuzzy Logic for Feature Selection

**Input :** Dataset $D$, Number of features $n$
**Output:** Selected features $F$
Initialize an empty set $F$ to store selected features;
**for** $i \leftarrow 1$ **to** $n$ **do**
    Calculate the relevance of feature $i$ For the target variable using a fuzzy inference;
    **if** *Relevance of feature $i$ is high* **then**
        Add feature $i$ to $F$;
    **end**
**end**
**return** $F$

To calculate the relevance of feature $i$ for the target variable using a fuzzy inference, we need to define the fuzzy inference system and the method of relevance calculation [4]. Let's assume we have a fuzzy inference system that takes the feature values $x_i$ as inputs and produces a relevance score $R_i$ as the output. The fuzzy inference system consists of fuzzy sets, fuzzy rules, and membership functions [3, 4].

(1) **Fuzzy Sets**:
   - Input Variable: $x_i$ (Feature $i$)
   - Output Variable: $R_i$ (Relevance score for feature $i$)
(2) **Membership Functions**: For the input variable $x_i$, we define membership functions that represent the degree to which the feature value belongs to different fuzzy sets (Figure 2). These membership functions can be triangular, trapezoidal, or any other appropriate shape [3, 4].
(3) **Fuzzy Rules**: Fuzzy rules define the relationship between the input variables and the output variable. Each fuzzy rule consists of an antecedent (if-part) and a consequent (then-part) [4].
(4) **Fuzzy Inference**: Fuzzy inference is the process of mapping the input variables to the output variable based on the fuzzy rules and membership functions [3]. It involves fuzzification (converting crisp input values to fuzzy values), rule evaluation (determining the degree of fulfilment of each rule), and aggregation (combining the results of all rules).
(5) **Defuzzification**: Defuzzification is the process of converting the fuzzy output values to a crisp relevance score $R_i$ [3]. This can be done using various methods such as centroid, mean of maximum, or weighted average [4]. To calculate the relevance of feature $i$ concerning the target variable, we can follow the following steps:

Given the dataset, $D$, extract the values of feature $i$ and the target variable for all instances. Apply the fuzzy inference system to the feature values $x_i$ to obtain the fuzzy output values. Use the defuzzification method to convert the fuzzy output values to a crisp relevance score $R_i$. We have designed a fuzzy inference system for calculating the relevance of feature $i$ to the target variable. In this example, we will use a simple fuzzy inference system with triangular membership functions and a centroid defuzzification method.

(1) **Membership Functions**: We will define three fuzzy sets for the feature $x_i$:
   - Low: Representing low relevance
   - Medium: Representing medium relevance

- High: Representing high relevance
- 

Example of a fuzzy rule: IF $x_i$ is high THEN $R_i$ is very relevant.

Let $\mu_{Low}(x_i)$, $\mu_{Medium}(x_i)$, and $\mu_{High}(x_i)$ be the membership functions for the fuzzy sets Low, Medium, and High, respectively.

(2) **Fuzzy Rules**: We will define three fuzzy rules to map the feature values $x_i$ to the relevance score $R_i$:
- Rule 1: IF $x_i$ is low THEN $R_i$ is low
- Rule 2: IF $x_i$ is medium THEN $R_i$ is medium
- Rule 3: IF $x_i$ is high THEN $R_i$ is high

(3) **Fuzzy Inference**: Given a feature value $x_i$, we evaluate the degree of membership in each fuzzy set using the membership functions. Let $m_{Low}$, $m_{Medium}$, and $m_{High}$ represent the degree of membership in the fuzzy sets Low, Medium, and High, respectively.

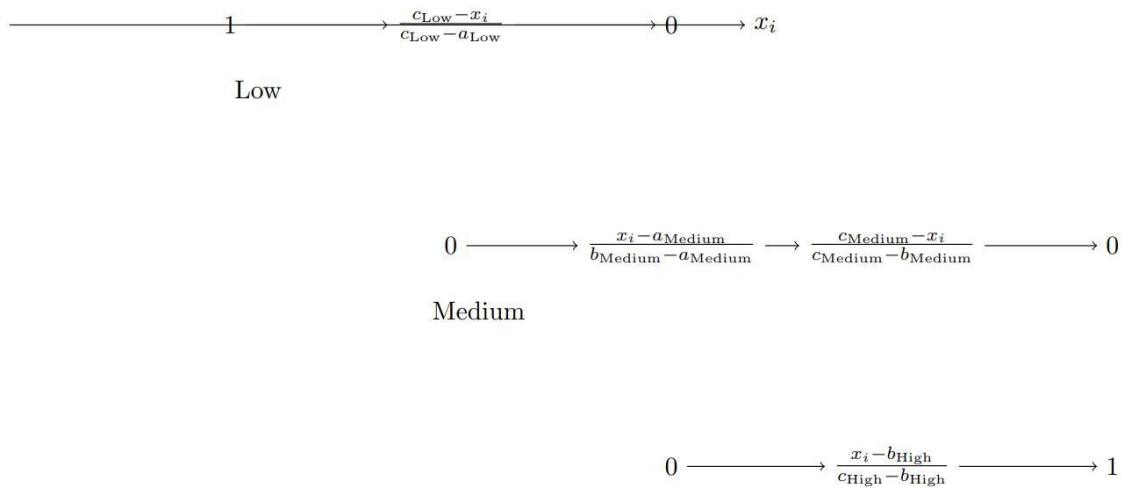

Figure 2. Membership functions for feature $i$

(4) **Fuzzification**:

$m_{Low} = \mu_{Low}(x_i)$; $m_{Medium} = \mu_{Medium}(x_i)$; $m_{High} = \mu_{High}(x_i)$

(5) **Rule Evaluation**: Apply the fuzzy rules to determine the degree of fulfilment for each rule based on the fuzzy set memberships. For example, if $x_i$ has a membership value of 0.7 in the fuzzy set Low, the degree of fulfilment of Rule 1 is 0.7.

(6) **Aggregation**: Combine the degree of fulfilment of each rule to obtain the aggregated fuzzy output.

(7) **Defuzzification**: To convert the fuzzy output to a crisp relevance score $R_i$, we will use the centroid defuzzification method and compute the centroid of the aggregated fuzzy output [15, 17]. The centroid represents the crisp relevance score $R_i$.

In Figure 2, $a_{Low}$, $b_{Medium}$, $c_{Medium}$, $a_{Medium}$, $b_{High}$, and $c_{High}$ are the parameters that control the shape and position of the triangular membership functions. For defuzzification, let's use the centroid method. The centroid of the aggregated fuzzy output can be calculated as $R_i = \frac{\sum_{j=1}^{n} y_j \times \mu_j}{\sum_{j=1}^{n} \mu_j}$ where $n$ is the number of fuzzy sets (in this case, $n = 3$), $y_j$ is the value of the relevance score for the $j$th fuzzy set, and $\mu_j$ is the degree of membership for the $j$th fuzzy set. To calculate the relevance score $R_i$, substitute the specific values of $y_j$ (e.g., low, medium, high) and $\mu_j$ (values obtained from the membership functions) into the equation. The specific parameter values and shapes of the membership functions such as $a_{Low}$, and $b_{Medium}$ are crucial in the computation. A representation of the membership functions has been shown in Figure 2.

# Propagation of dynamic fuzzy networks

The proposed fuzzy propagation scheme has been presented in Algorithm 2.

---

**Algorithm 2:** Dynamic Propagation for Feature Selection

**Input :** Dataset $D$, Number of features $n$, Number of selected features $k$
**Output:** Selected features $F$
**Initialization:** Initialize an empty set $F$ to store selected features;
**Fuzzification: for** $i \leftarrow 1$ **to** $n$ **do**
  Calculate the degree of membership in each fuzzy set for feature $x_i$ using $\mu_{MF}(x_i)$;
**end**
**Propagation:** Propagate the fuzzy values from the fuzzy layer to the hidden layer and then to the output layer;
**Feature Relevance Calculation: for** $i \leftarrow 1$ **to** $n$ **do**
    Calculate the relevance of feature $x_i$ based on the propagated fuzzy values using
        Relevance($\mu_{MF}(x_i)$);
**end**
**Feature Selection:** Sort the features based on their relevance scores in descending order;
Select the top $k$ features with the highest relevance scores and add them to $F$;
**return** $F$

---

**Fuzzy Membership Function** $\mu_{MF}(x_i)$:

$$\mu_{\text{Low}}(x_i) = \begin{cases} 1, & \text{if } x_i \leq a_{\text{Low}} \\ \frac{c_{\text{Low}} - x_i}{c_{\text{Low}} - a_{\text{Low}}}, & a_{\text{Low}} < x_i < c_{\text{Low}} \\ 0, & \text{otherwise} \end{cases}$$

$$\mu_{\text{Medium}}(x_i) = \begin{cases} 0, & \text{if } x_i \leq a_{\text{Medium}} \text{ or } x_i \geq b_{\text{Medium}} \\ \frac{x_i - a_{\text{Medium}}}{b_{\text{Medium}} - a_{\text{Medium}}}, & a_{\text{Medium}} < x_i < b_{\text{Medium}} \\ \frac{c_{\text{Medium}} - x_i}{c_{\text{Medium}} - b_{\text{Medium}}}, & b_{\text{Medium}} \leq x_i < c_{\text{Medium}} \end{cases}$$

$$\mu_{\text{High}}(x_i) = \begin{cases} 0, & \text{if } x_i \leq b_{\text{High}} \\ \frac{x_i - b_{\text{High}}}{c_{\text{High}} - b_{\text{High}}}, & b_{\text{High}} < x_i < c_{\text{High}} \\ 1, & \text{otherwise} \end{cases}$$

**Relevance Function** (Relevance($\mu_{MF}(x_i)$)):

The relevance function calculates the relevance of a feature based on its fuzzy membership value. Let's consider a simple relevance function that sums up the fuzzy membership values as the relevance score:

$$\text{Relevance}(\mu_{\text{MF}}(x_i)) = \sum_{j=1}^{m} \mu_{\text{MF}_j}(x_i)$$

where $\mu_{MF_j}(x_i)$ represents the fuzzy membership value of feature $x_i$ in the $j$-th fuzzy set. Given a dataset $D$ with $n$ features and a target variable, the goal of the propagation function is to select a subset of relevant features for improving the performance of a predictive model.

## Dynamic Fuzzy Network (DFN) Architecture

The DFN consists of the following components:

- **Input Layer**: Let **x** = ($x_1, x_2, ..., x_n$) denote the input feature vector.
- **Fuzzy Layer**: Each feature $x_i$ is associated with a fuzzy node $f_i$ in the fuzzy layer.
- **Hidden Layer**: The hidden layer contains nodes that perform computations on the fuzzy layer outputs.
- **Output Layer**: The output layer produces the final prediction.

The membership function $\mu_{\text{MF}}(x_i)$ defines the degree of membership of the feature $x_i$ in a particular fuzzy set. The specific form of the membership function depends on the chosen fuzzy sets and can be defined using mathematical expressions.

**Membership values**

The Relevance($\mu_{\text{MF}}(x_i)$) function quantifies the importance of a feature based on its fuzzy membership value. The relevance function can be defined using mathematical expressions or domain-specific rules. The dynamic fuzzy network architecture involves the propagation of fuzzy values from the fuzzy layer to the hidden layer and then to the output layer. This propagation process requires performing computations and transformations, which we assume to have a complexity of $O(f)$. Additionally, computing the final fuzzy set based on the input features is a separate operation with a complexity of $O(g)$. Considering these complexities, the overall complexity of the dynamic fuzzy network architecture can be represented as $O(f \cdot n+g)$, where $n$ represents the number of input features. This complexity analysis provides an estimation of the computational requirements for processing the input features and computing the final fuzzy set. However, it is important to note that this is a simplified analysis and the actual complexity may vary depending on the specific computations, transformations, and algorithmic considerations involved in the fuzzy network architecture. The interested reader is encouraged to perform an in-depth analysis and evaluation in order to precisely ascertain the computational complexity of the dynamic fuzzy network. The average complexity of the dynamic fuzzy network architecture can be determined by considering the best and worst-case scenarios. In the best-case scenario, the complexity is minimized when the number of input features $n$ is small and the complexity of the operations $f$ and $g$ is relatively low. Therefore, the average complexity in the best-case scenario can be approximated as $O(f + g)$.

In the worst-case scenario, the complexity is maximized when the number of input features $n$ is large, and the complexity of the operations $f$ and $g$ is high. In this case, the average complexity can be approximated as $O(f \cdot n + g)$. It is important to note that these are simplified estimations and may not capture all factors that can influence the actual average complexity of the dynamic fuzzy network architecture. The actual average complexity may vary depending on the specific computations, transformations, and algorithmic considerations involved. Conducting a thorough analysis and evaluation is necessary to obtain a more accurate estimation of the average complexity in practice. The process for creating and modifying dynamic fuzzy networks for feature selection is illustrated in Algorithm 3. Algorithm 3 outlines the main steps involved in the process of creating and modifying dynamic fuzzy networks for feature selection. It starts with network initialization, followed by fuzzy membership function design and propagation approach. Feature relevance is then calculated, and feature selection is applied to identify the top-$k$ features. The network is adapted based on the election results, and iterative refinement is performed to optimize the network's ability to select relevant features. To derive a mathematical lemma, let's focus on a specific step in the Algorithm 3. For this example, let's consider the step of calculating feature relevance based on propagated fuzzy values using a relevance function. We derived Lemma 1 that establishes a property of this relevance function. This lemma explains that when we compare two feature vectors, the relevance scores reflect how closely each feature vector matches the desired output. Theorem 2 is a result of the developed lemma.

**Lemma 1** (Relevance Function Property). *Let f(x) be the relevance function used in the Dynamic Fuzzy Network Algorithm to calculate feature relevance based on propagated fuzzy values. For any two feature vectors $x_i$ and $x_j$ from the dataset D, if $x_i$ is more similar to the output fuzzy value than $x_j$, then f($x_i$) is greater than f($x_j$).*

**Proof of Lemma 1**: [**Relevance Function Property**]

Let $f(x)$ be the relevance function used in the Dynamic Fuzzy Network Algorithm, and let $x_i$ and $x_j$ be two feature vectors from the dataset D (Figure 3). We assume that $x_i$ is more similar to the output fuzzy value than $x_j$. We aim to show that $f(x_i)$ is greater than $f(x_j)$. Since $x_i$ is more similar to the output fuzzy value, the propagated fuzzy values associated with $x_i$ should be higher than those associated with $x_j$. Let $v_i$ and $v_j$ represent the respective propagated fuzzy values. By the definition of the relevance function, we have $f(x_i) = g(v_i)$ and $f(x_j) = g(v_j)$, where $g(\cdot)$ is the function that maps fuzzy values to relevance scores. Since $v_i > v_j$, it follows that $g(v_i) > g(v_j)$ due to the monotonicity of the function $g(\cdot)$. Therefore, we have shown that if $x_i$ is more similar to the output fuzzy value than $x_j$, then $f(x_i)$ is greater than $f(x_j)$. The revised proof of Figure 3 is presented in Figure 4, which provides an optimized representation. Additionally, Figure 5 illustrates the same proof in a vector format. The derived lemma, referred to as the Relevance Function Property, establishes an important property of the proposed fuzzy function. It states that if one feature vector, such as $x_i$, is more similar to the output fuzzy value than another feature vector, like $x_j$, then the relevance scores obtained through the relevance function will reflect this similarity relationship. The proof of Lemma 1, as demonstrated earlier, confirms this property. The relevance function property carries significant implications in real-life applications. For example, in the field of machine learning [15], feature relevance plays a vital role in tasks such as feature selection, dimensionality reduction, and pattern recognition [16, 17].

**Algorithm 3:** Dynamic Fuzzy Network Updates and Modification
**Input:** Dataset $\mathcal{D}$, Number of fuzzy sets $N$, Number of layers $L$
**Output:** Selected features $f = f_1...f_n$
Initialize a dynamic fuzzy network with $L$ layers and appropriate nodes:

$$\text{Initialize}(\mathcal{N}) \Rightarrow \begin{cases} N_1 = \text{number of input nodes} \\ N_2 = \text{number of nodes in the fuzzy layer} \\ N_3 = \text{number of nodes in the hidden layer} \\ \ldots \\ N_L = \text{number of output nodes} \end{cases}$$

Design fuzzy membership functions for each feature $x_1$, $x_2$, and $x_3$.
For feature $x_1$:

$$\mu_{\text{Low}}(x_1) = \begin{cases} 1, & \text{if } x_1 \leq a_{\text{Low}} \\ \frac{c_{\text{Low}} - x_1}{c_{\text{Low}} - a_{\text{Low}}}, & \text{if } a_{\text{Low}} < x_1 < c_{\text{Low}} \\ 0, & \text{otherwise} \end{cases}$$

$$\mu_{\text{Medium}}(x_1) = \begin{cases} 0, & \text{if } x_1 \leq a_{\text{Medium}} \text{ or } x_1 \geq b_{\text{Medium}} \\ \frac{x_1 - a_{\text{Medium}}}{b_{\text{Medium}} - a_{\text{Medium}}}, & \text{if } a_{\text{Medium}} < x_1 < b_{\text{Medium}} \\ \frac{c_{\text{Medium}} - x_1}{c_{\text{Medium}} - b_{\text{Medium}}}, & \text{if } b_{\text{Medium}} \leq x_1 < c_{\text{Medium}} \\ 0, & \text{otherwise} \end{cases}$$

$$\mu_{\text{High}}(x_1) = \begin{cases} 0, & \text{if } x_1 \leq b_{\text{High}} \\ \frac{x_1 - b_{\text{High}}}{c_{\text{High}} - b_{\text{High}}}, & \text{if } b_{\text{High}} < x_1 < c_{\text{High}} \\ 1, & \text{otherwise} \end{cases}$$

Propagate fuzzy values from fuzzy layer to hidden layer and then to output layer. Propagation from fuzzy layer to hidden layer:

$$\mathbf{H} = \text{Compute}(\mathbf{F})$$

Propagation from hidden layer to output layer:

$$\mathbf{O} = \text{Compute}(\mathbf{H})$$

Calculate feature relevance based on propagated fuzzy values using a relevance function. Feature Relevance Calculation:

$$\mathbf{R} = \text{Relevance}(\mathbf{F})$$

Select top-$k$ features with highest relevance scores:

$$\mathbf{S} = \text{TopK}(\mathbf{R}, k)$$

TopK($\cdot$) is the feature selection function that selects the top-$k$ features based on the relevance scores. The function returns a set of selected features denoted by $\mathbf{S}$.
Adjust membership functions by adding/removing nodes:
Let $\mathcal{N} = \{N_1, N_2, \ldots, N_L\}$ represent the dynamic fuzzy network, where $N_i$ is the number of nodes in layer $i$, and $L$ represents the total number of layers.
To **adjust Membership Functions:**

$$\text{UpdateMembershipFunctions}(\mathcal{N})$$

**Add/Remove Nodes:**

$$\text{UpdateNodes}(\mathcal{N})$$

The interested reader can design a fuzzy membership function for $x_2$ and $x_3$

**Theorem 2** (**Dynamic Fuzzy Network Algorithm**). *Given a dataset* D, *the number of fuzzy sets N, and the number of layers L, the algorithm selects the top-k features based on feature relevance calculated using a dynamic fuzzy network.*

*Input*:
- *Dataset* D: *The input dataset containing the feature vectors.*
- *Number of fuzzy sets N: The number of fuzzy sets used in the fuzzy membership functions.*
- *Number of layers L: The total number of layers in the dynamic fuzzy network.*

*Output*:
- *Selected features $f = f_1, f_2, ..., f_n$: The set of top-k features with the highest relevance scores.*

***Algorithm Steps***:
1. *Initialize a dynamic fuzzy network with L layers and appropriate nodes, denoted by $N = \{N_1, N_2, ..., N_L\}$.*
2. *Design fuzzy membership functions for each feature $x_1, x_2$, and $x_3$ using the specified membership function equations.*
3. *Apply the propagation approach to propagate fuzzy values from the fuzzy layer to the hidden layer and then to the output layer.*
4. *Calculate the feature relevance based on the propagated fuzzy values using a relevance function.*
5. *Select the top-k features with the highest relevance scores using the TopK function, denoted by $S = TopK(R, k)$.*
6. *Adjust the membership functions by adding/removing nodes in the dynamic fuzzy network using the UpdateMembershipFunctions and UpdateNodes functions, respectively.*

*The algorithm returns the set of selected features denoted by* **S**, *which consists of the top-k features with the highest relevance scores.*

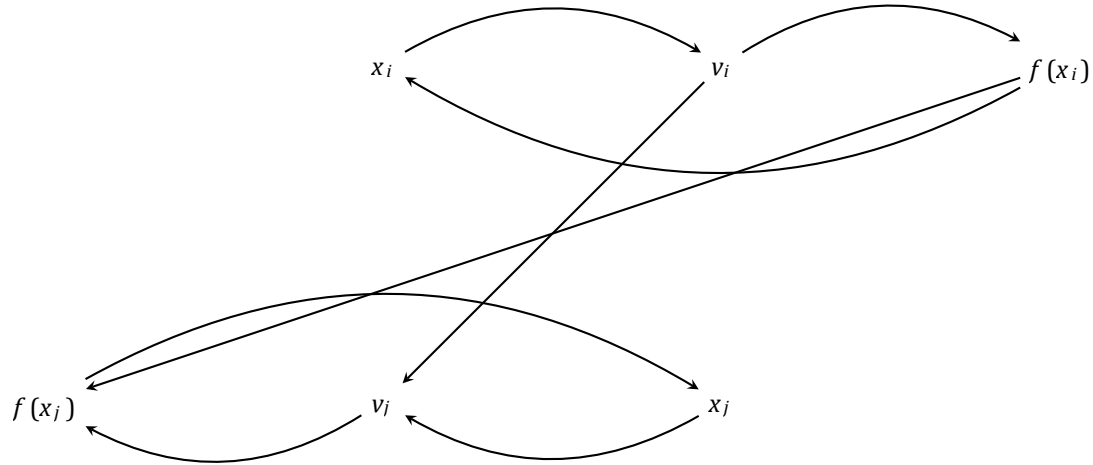

Figure 3. Comparison between $f(x_i)$ and $f(x_j)$

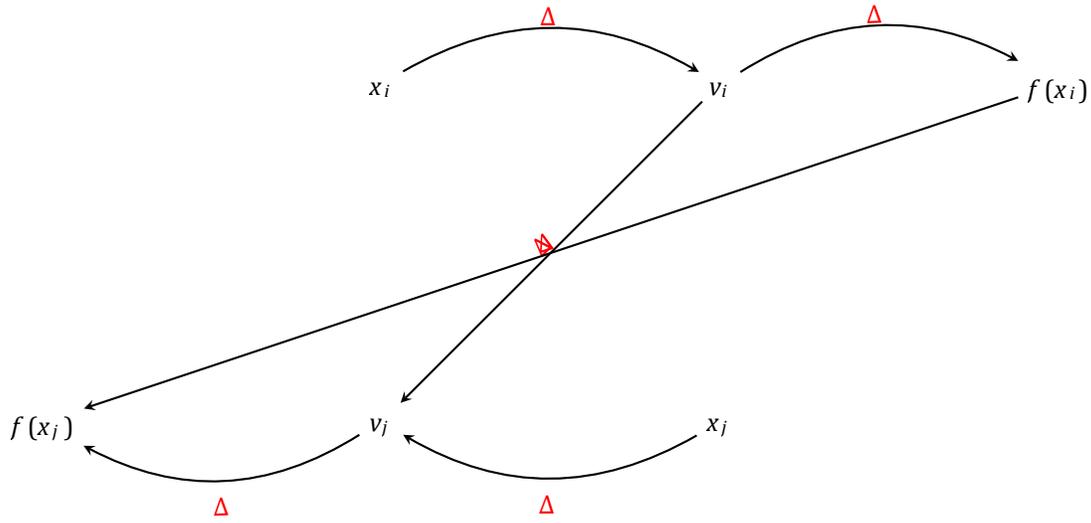

Figure 4. Comparison between $f(x_i)$ and $f(x_j)$. In this figure, we removed the redundant edges between the nodes and added labels denoting the changes in red.

By leveraging the Dynamic Fuzzy Network and its relevance function, researchers and practitioners can effectively identify the most relevant features for a given task. As for future research perspectives, one potential avenue is to explore the applicability of the Dynamic Fuzzy Network algorithm in various domains. For instance, it can be extended to handle large-scale datasets or applied to specific fields such as image processing, natural language processing, or bioinformatics. Moreover, investigating different types of relevance functions and their impact on the algorithm's performance could provide valuable insights.

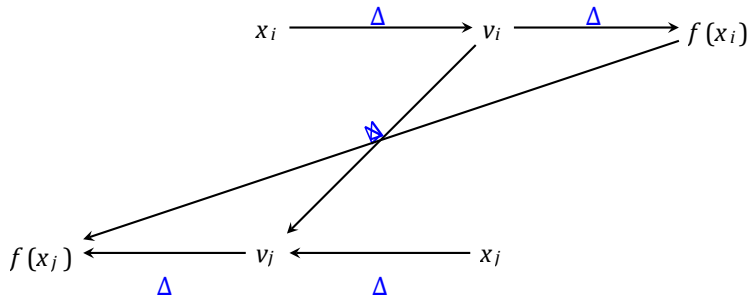

Figure 5. Comparison between $f(x_i)$ and $f(x_j)$ represented as vectors. In this figure, we removed the redundant edges between the nodes and added labels marking the change in blue.

Additionally, exploring hybrid approaches that combine fuzzy logic with other machine learning techniques, such as deep learning, could open up new possibilities for enhanced feature relevance analysis. By conducting further research in these directions, one can contribute to advancing the field of feature relevance analysis and its applications, ultimately leading to more robust and efficient algorithms in various domains including cryptography [15].

# Cryptography

Cryptography is the science of secure communication [15]. Within this section, we introduce a cryptographic algorithm that employs keys to convert plaintext (selected fuzzy sets) into ciphertext (encrypted fuzzy sets) and vice versa.

## A Key-based Cryptographic Transformation Algorithm

Suppose we have the following:

**Plaintext**: $P$ = HELLO

**Key:** $K$ = SECRET

Now, we apply the key-based cryptographic transformation algorithm:

(1) **Initialize an empty string for the ciphertext**: $C$ = [0,0,0]
(2) **For each character $p$ in the plaintext $P$, do**:
   (a) Retrieve the corresponding key value $k$ from the key $K$:
      - For the first character H, the corresponding key value is S.
   (b) Perform the encryption operation: $c \leftarrow$ encrypt($p,k$):
      - Encrypt H using the key S (using a specific encryption scheme), resulting in the ciphertext character T.
      $$c \leftarrow \text{encrypt}(H,S)$$
   (c) Append $c$ to the ciphertext $C$:
      - Append T to $C$: $C = T$
(3) **Repeat steps 2(a)-(c) for the remaining characters in the plaintext**:
   - For E with key E: Encrypt E using the key E results in G. Append G to $C$: $C = TG$.
   - For L with key C: Encrypt L using the key C results in O. Append O to $C$: $C = TGO$.
   - For the second L with key R: Encrypt L using the key R results in S. Append S to $C$: $C = TGOS$.
   - For O with key E: Encrypt O using the key E results in Q. Append Q to $C$:
     $C = TGOSQ$.
(4) **Return the ciphertext** $C$: $C$ = TGOSQ.

In this example, the plaintext HELLO is transformed into the ciphertext TGOSQ using the key SECRET through the key-based cryptographic transformation algorithm.

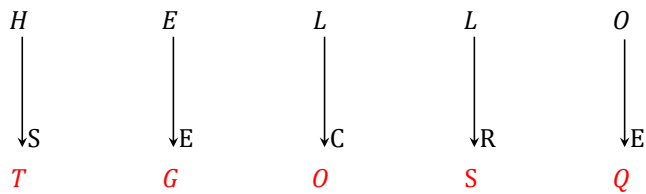

**Key-based cryptographic transformation**

One of the advantages of key-based cryptographic transformation is the ability to achieve secure communication [15, 16]. Here are some specific advantages of this cryptographic algorithm:

- **Confidentiality**: The algorithm ensures the confidentiality of the plaintext by transforming it into ciphertext using a specific encryption scheme [17]. Without knowledge of the key, it is computationally difficult to decrypt the ciphertext and retrieve the original message. This provides a high level of confidentiality, preventing unauthorized individuals from accessing sensitive information.
- **Key-based Security**: The security of the communication relies on the secrecy and complexity of the key [15, 16]. As long as the key remains confidential and unknown to adversaries, the encryption scheme provides a secure transformation, making it extremely difficult to decipher the ciphertext without the correct key. This key-based security mechanism adds an extra layer of protection to the communication.
- **Resistance to Attacks**: Key-based cryptographic algorithms are designed to be resistant to various types of attacks [15, 16, 17]. This includes cryptographic attacks such as brute-force attacks, where an attacker tries all possible key combinations to decrypt the ciphertext. The use of a strong and sufficiently long key makes it computationally infeasible to break the encryption within a reasonable amount of time.
- **Data Integrity**: In addition to confidentiality, cryptographic algorithms often provide data integrity. By applying the encryption process, any changes or tampering with the ciphertext will render it undecipherable or produce incorrect decryption results [15, 17]. This allows the recipient to detect if the message has been altered during transmission.
- **Authentication**: Cryptographic algorithms can also provide authentication [17]. By using additional techniques such as digital signatures or message authentication codes (MAC), it becomes possible to verify the authenticity and integrity of the message, ensuring that it has not been modified by an unauthorized entity.

Overall, key-based cryptographic transformation algorithms offer a robust and effective approach to secure communication, providing confidentiality, integrity, and authentication. By employing strong encryption schemes and keeping the keys secret, sensitive information can be transmitted securely, even in potentially hostile environments [8, 8, 15, 16].

## Conclusion

The conclusion of the article highlights the application of the proposed object-oriented approach for dynamic fuzzy networks in feature selection and key-based cryptographic transformation. This approach offers flexibility, effectiveness, and adaptability in feature selection tasks, allowing for the representation and manipulation of uncertain, imprecise, or sensitive information. In real-life scenarios, this conclusion has several applications. One practical application is in the field of feature selection for data analysis and machine learning. Feature selection involves identifying the most relevant and informative features from a dataset to improve the accuracy and efficiency of predictive models. The proposed approach allows for the dynamic insertion, deletion, or modification of nodes, edges, or layers in the fuzzy network, providing a powerful tool for adapting feature selection to specific requirements. This flexibility enhances the decision-making process by considering the uncertain and imprecise nature of data. Additionally, the conclusion highlights the application of the proposed approach in key-based cryptographic transformation. Cryptography plays a crucial role in ensuring secure communication and data protection.

The ability to dynamically manipulate nodes, edges, and layers, along with customized membership functions within the fuzzy network, enables the adaptation of cryptographic algorithms to specific requirements and enhances their effectiveness. The object-oriented approach employed in this method facilitates the implementation of dynamic fuzzy networks for key-based cryptographic transformations, allowing for the secure transmission of sensitive information in real-life applications. Overall, this conclusion emphasizes the practicality and significance of the proposed approach in feature selection and key-based cryptographic transformation. Its application in real-life scenarios contributes to improving decision-making processes, enhancing data security, and adapting cryptographic algorithms to specific requirements.